\documentclass{iaus}
\usepackage{graphicx}

\title[JD 11.~~Pre-stellar cores and IRDCs] %% give here short title %%
{Observational studies of pre-stellar cores and infrared dark clouds}

\author[P. Caselli]   %% give here short author list %%
{Paola Caselli$^{1}$}
\affiliation{$^1$School of Physics and Astronomy, University of Leeds, Leeds LS2 9JT, UK 
 \\email: {\tt p.caselli@leeds.ac.uk}} 

\pubyear{2011}
\volume{280}  %% insert here IAU Symposium No.
\pagerange{1-13}
\jname{The Molecular Universe}
\editors{J. Cernicharo \& R. Bachiller, eds.}
\begin{document}

\maketitle

\begin{abstract}
Stars like our Sun and planets like our Earth form in dense regions within interstellar molecular clouds, called pre-stellar cores (PSCs). PSCs provide the initial conditions in the process of star and planet formation. In the past 15 years, detailed observations of (low-mass) PSCs in nearby molecular cloud complexes have allowed us to find that they are cold (T $<$ 10\,K) and quiescent (molecular line widths are close to thermal), with a chemistry profoundly affected by molecular freeze-out onto dust grains. In these conditions, deuterated molecules flourish, becoming the best tools to unveil the PSC physical and chemical structure. Despite their apparent simplicity, PSCs still offer puzzles to solve and they are far from being completely understood. For example, what is happening to the gas and dust in their nuclei (the future stellar cradles) is still a mystery that awaits for ALMA. Other important questions are: how do different environments and external conditions affect the PSC physical/chemical structure? Are PSCs in high-mass star forming regions similar to the well-known low-mass PSCs? Here I review observational and theoretical work on PSCs in nearby molecular cloud complexes and the ongoing search and study of massive PSCs embedded in infrared dark clouds (IRDCs), which host the initial conditions for stellar cluster and high-mass star formation.
\end{abstract}

\firstsection % if your document starts with a section,
\section{Introduction}
For about 30 years we have known that stars like our Sun form within the dense regions of interstellar molecular clouds (e.g. Benson \& Myers 1989; Beichman et al. 1986), where submicron sized dust grains and molecules such as H$_2$, CO and more complex organic species coexist and interact (Herbst \& van Dishoeck 2009). These regions are called dense cores (Myers et al. 1983) and represent the initial conditions of the star formation process (e.g. Shu et al. 1987). In fact, observations in quiescent regions, such as the Pipe nebula (Alves et al. 2007; Rathborne et al. 2009) and the Aquila Rift (Andr\'e et al. 2010; K{\"o}nyves et al. 2010) indicate that the stellar mass function is the reflection of the core mass function (CMF). If so, processes controlling the formation of the cores sets the initial mass function (IMF; Salpeter 1955). Observations of dense cores and their interpretation are therefore fundamental to put constraints on star and planet formation theory as well as on simulations of molecular cloud formation and evolution.  How can we study the initial conditions of star and planet formation by looking at dense cores?   

The initial conditions are soon lost as dense cores become associated with protostars, which drive powerful outflows and alter the physical and chemical properties of the parent cloud (e.g. Arce et al. 2007, 2010; Foster et al. 2009; Friesen et al. 2010a; Pineda et al. 2011). Thus one first needs to identify dense cores on the verge of star formation, the so-called pre-stellar cores (PSCs). PSCs are typically cold (temperature  $\le$10\,K) and have volume densities $>$\,$\simeq$2$\times$10$^{-19}$\,g\,cm$^{-3}$, or number densities $>$5$\times$10$^{4}$\,H$_2$ molecules per cc (e.g. Ward-Thompson et al. 1994; Crapsi et al. 2005).  In such environments, the dust mostly emits sub-millimetre radiation (peaking around 300$\mu$m), and the rotational transitions of molecules such as CO emit in the readily accessible millimetre spectrum. While CO is the most abundant molecule after H$_2$, there are a host of other observable molecules that provide additional information. For example, ammonia (NH$_3$; e.g. Goodman et al. 1993) and dyazenilium (N$_2$H$^+$; Caselli et al. 1995, 2002a) both have lower optical depths and are better tracers of dense gas. Both molecules have hyperfine spectra useful in determining the gas temperature and column densities.  Over the past 15 years, observations of dust, CO, NH$_3$, and N$_2$H$^+$ have established the basic structure of the cores: nearly hydrostatic equilibrium and heated from the outside by starlight. More recent studies have been able to analyze observations within this basic framework to deduce some of the physical and chemical processes in molecular clouds such as internal wave motions, UV heating, molecular depletion and desorption between gas and solid phases, and chemical fractionation. In Sect.\,\ref{sec_lowmass} below we discuss this further. The more recent study of PSCs in high-mass star forming regions is described in Sect.\,\ref{sec_highmass}.

\section{Low-mass PSCs}
\label{sec_lowmass}

The state of the art in the PSCs field is summarized below. More comprehensive reviews can be found in Di Francesco et al. (2007) and Bergin \& Tafalla (2007). 

\subsection{Molecular freeze-out}
First identified by Ward-Thompson et al. (1994) as centrally concentrated cores in the dust millimetre continuum emission, PSCs are characterized by relatively large central number densities (a few $\times$\,10$^6$\,cm$^{-3}$), low central kinetic temperature (T$\simeq$6\,K; Crapsi et al. 2007), and kinematic features which can be reproduced by gravitational collapse motions  (e.g. Caselli et al. 2002b; van der Tak et al. 2005). CO molecules (typically used to derive cloud masses, assuming a ÒcanonicalÓ CO/H$_2$ abundance ratio of $\simeq$10$^{-4}$, as measured in nearby molecular clouds;  e.g. Frerking et al. 1987, Pineda et al. 2008) do not trace the dust peak at all (Willacy et al. 1998; Caselli et al. 1999; Bacmann et al. 2002; Pagani et al. 2007). To reproduce observations toward the PSC L1544, about 90\% of CO molecules should leave the gas phase, on average along the line of sight, and over 99\% of them must deplete in the core nucleus (Caselli et al. 1999). The only easy way to get rid of gas-phase molecules in quiescent and cold regions is to deposit them on the surface of dust particles, forming thick icy mantles (e.g. Leger 1983; Lacy et al. 1994; Ossenkopf \& Henning 1994; Ormel et al. 2009; Chiar et al. 2011). 

Unlike CO and other C-bearing species (Tafalla et al. 2006), N-bearing molecules do not significantly freeze-out in PSC centers (Hily-Blant et al. 2010): only a factor $\simeq$2 abundance drop for N$_2$H$^+$ (Caselli et al. 2002c, Bergin et al. 2002) and hints of abundance rise for NH$_3$ (Tafalla et al. 2004). This is a puzzle in regions where CO is heavily depleted, given that the binding energy of N$_2$ (the precursor of N$_2$H$^+$ and NH$_3$) onto bare grains and icy surfaces is similar to the CO binding energy (\"Oberg et al. 2005) and the sticking coefficient is the same for the two species ($\simeq$1; Bisschop et al. 2006). Also, NH$_3$ is significantly more strongly bound on grain mantles than CO. A possible explanation may be the slow conversion of atomic nitrogen into N$_2$, because of the slow neutral-neutral reactions involved in the process (Hily-Blant et al. 2010), the possibly low sticking probability of N ($\simeq$0.1; Flower et al. 2006) and the fact that the freeze-out of CO decreases the destruction rate of N$_2$H$^+$ and molecular ions important for the formation of NH$_3$ (such as NH$^+$; see also Aikawa et al. 2005, 2008). Indeed, Maret et al. (2006) found evidence of a low fraction of N$_2$ ($\simeq$10$^{-6}$ with respect to H nuclei) and a consequent high N/N$_2$ ratio toward the starless core B68, although some N has to be trapped on the surface of dust grains to avoid a too efficient production of N$_2$.  Interestingly, Hily-Blant et al. (2010) also require a low fractional abundance of atomic nitrogen in the gas phase  ($\le$ 10$^{-6}$), to reproduce the observed abundance ratio of N-bearing species tracing the CO-depleted zone of five starless cores (but uncertainties on the rate coefficients of key neutral-neutral reactions prevent unambiguous conclusions).   In any case, in the central few hundred AU of a PSC (the PSC nucleus), where the volume density is $>$10$^6$\,cm$^{-3}$, molecules freeze-out onto dust grains with time scales $\le$10$^3$/[$n({\rm H_2})/10^6$\,cm$^{-3}$]\,yr (assuming a gas-to-dust mass ratio of 100 and a dust grain radius of 0.1\,$\mu$m; e.g. van Dishoeck et al. 1993, Flower et al. 2005), and no high density tracer is expected to survive in the gas phase (except for those molecules composed by the volatile light elements H and D, such as H$_3^+$ and its deuterated isotopologues; see Sect.\,\ref{sub_deuterium}). An example of this is provided by the recent interferometric observations of N$_2$H$^+$(1-0) toward Oph\,B (Friesen et al. 2010b), where no emission was found toward dust continuum peaks within the core.  

Keto \& Caselli (2008, 2010) used the basic temperature-density model to predict line intensities and profiles for different values of the cosmic ray ionization rate and the reverse process to depletion, molecular desorption, which is dependent on cosmic rays. Detailed comparison with observed molecular lines put limits on the rate of cosmic-ray ionization ($\simeq$1$\times$10$^{-17}$\,s$^{-1}$) and desorption (about 30 times larger than the value deduced by Hasegawa \& Herbst 1993 and in agreement with more recent studies by Bringa \& Johnson 2004 and Herbst \& Cuppen 2006).  Moreover, to reproduce the low temperature observed toward the core nucleus ($\simeq$6\,K) the dust opacity has to increase by a factor of about 4 compared to the standard dust opacities of Ossenkopf \& Henning (1994).  This is consistent with fluffy dust, suggesting that dust grains have started to coagulate (see also Ormel et al. 2009), with a consequent reduction of the molecular freeze-out rate because of the smaller surface area available for adsorption (Flower et al. 2005). 

\subsection{Deuterium fractionation}
\label{sub_deuterium}

Deuterated molecules are highly enhanced in PSC nuclei.  The reason for this is briefly outlined here. In cold and dark clouds (such as PSCs), H$_2$ molecules are mainly destroyed by cosmic rays, which produce H$_2^+$.  As soon as H$_2^+$ is formed, it reacts with another H$_2$ molecule, forming one of the most important molecular ions in astrochemistry: H$_3^+$. This molecular ion can easily cede a proton to atoms and other species, starting astrochemical processes and increasing the molecular complexity in the interstellar medium.  H$_3^+$ can also exchange its proton with a deuteron, through the reaction: H$_3^+$ + HD $\rightarrow$ H$_2$D$^+$ + H$_2$ + $\Delta$E ($\Delta$E = 230\,K; Millar et al. 1989). If the gas temperature is below 20\,K, such as in PSCs, the above exothermic reaction can only proceed from left to right, enhancing the H$_2$D$^+$/ H$_3^+$ abundance ratio.  This is reflected in species (such as DCO$^+$ and N$_2$D$^+$), which form in reactions of neutrals (CO and N$_2$, respectively) with H$_2$D$^+$. The ratio between deuterated and non-deuterated species (e.g. N$_2$D$^+$/ N$_2$H$^+$) is called D-fraction ($R_{\rm D}$) and its values can be as large as a few \%, about 3 orders of magnitude larger than the cosmic D/H ratio ($\simeq$10$^{-5}$; e.g. Linsky et al. 2006). 

$R_{\rm D}$ increases even farther (to about 13 orders of magnitude in multiply deuterated species; e.g. Ceccarelli et al. 2007) if the abundance of the main destruction partners of H$_3^+$ and H$_2$D$^+$ (mainly CO and O) is reduced, e.g. via freeze-out.  This was predicted by Dalgarno \& Lepp (1984) and observationally proved toward several starless cores and PSCs (Caselli et al. 2002c; Bacmann et al. 2003). Crapsi et al. (2005) found criteria to identify PSCs among starless cores by measuring $R_{\rm D}$ and studying line profiles: PSCs have (i) high $R_{\rm D}$ ($>$ 10\%); (ii) large amount of CO freeze-out ($>$ 99\%); (iii) large central density ($>$5$\times$10$^5$\,cm$^{-3}$) and centrally concentrated density profile; (iv) kinematics consistent with central gravitational infall. This is important because PSCs, the future stellar cradles, are the best candidates where to find the initial conditions in the process of star formation. Starless cores with lower densities and shallower density profiles may oscillate around an equilibrium configuration (e.g. Lada et al. 2003) or expand and disperse. 

The large amount of molecular freeze-out and the consequent boost in the D-fraction explain why PSCs are the strongest emitters of H$_2$D$^+$ and D$_2$H$^+$ in the Galaxy (Caselli et al. 2003, 2008; Vastel et al. 2004).  To reproduce these observations, chemical models had to include multiply deuterated species (Roberts et al. 2003). Being bright (main beam brightness temperatures $\simeq$1\,K), the  ortho-H$_2$D$^+$(1$_{10}$-1$_{11}$) line is an ideal tracer of PSC nuclei, even in those regions where most of the heavier species are gone from the gas phase (the "complete depletion" zone; Walmsley et al. 2004).    The emission of ortho-H$_2$D$^+$ extends over sizes $\simeq$5,000\,AU toward L1544 (Vastel et al. 2006), similar to the para-D$_2$H$^+$(1$_{10}$-1$_{01}$) line (Parise et al. 2011).  This is the size of the "deuteration zone", where CO is mostly frozen onto dust grains and bright lines of N$_2$D$^+$ and deuterated ammonia can be observed (e.g. Crapsi et al. 2005, 2007). Caselli et al. (2008) extended the search of ortho-H$_2$D$^+$ in a sample of PSCs and cores with embedded protostars.  Variations in the amount of $R_{\rm D}$ and ortho-H$_2$D$^+$ column densities were correlated with physical parameters and found that large uncertainties in the interpretation were present, due to our poor knowledge of the cosmic-ray ionization rate (Dalgarno 2006), the dust grain size distribution (Ormel et al. 2009), the H$_2$ ortho-to-para ratio (Troscompt et al. 2009; Pagani et al. 2009) and the abundance of atomic oxygen in molecular clouds and dense cores (Melnick \& Bergin 2005). Caselli et al. (2008) suggested that an important source of the scatter observed in the correlation plots is probably due to variations in the ortho-to-para ratio of H$_2$D$^+$, from $>$0.5 for PSCs, with gas temperatures  $<$10\,K, to 0.03 for protostellar cores, consistent with chemical models of dense cores with temperature $ \simeq$15\,K gas (Flower et al. 2004; see also Emprechtinger et al. 2009 and Friesen et al. 2010a for further discussion on the variation of deuterium fractionation toward protostellar objects).  

\subsection{Physical Structure}

Measurements of dust and gas temperature profiles (Crapsi et al. 2007; Pagani et al. 2007), as well as estimates of their density structure, kinematics, chemical composition and ionization fraction (e.g. Caselli et al. 2002b; Bergin et al. 2002; Schnee et al. 2007; Keto \& Caselli 2008, 2010) provided the following conclusions: (i) PSCs are mainly thermally supported (with some additional energy furnished by waves and subsonic turbulence; e.g. Caselli et al. 2002a; Lada et al. 2008). (ii) Subsonic oscillations and large scale wave motions are important to PSC internal energy (Keto \& Field 2005; Broderick et al. 2008); (iii) PCSs are heated from outside by starlight and throughout by cosmic rays (e.g. Evans et al. 2001; Zucconi et al. 2001); (iv) main coolants are molecular line radiation (in particular CO) in the outer regions and dust in the central nuclei (e.g. Goldsmith 2001; Galli et al. 2002); (v) their density structure is set by the Lane-Emden equation (e.g. Alves et al. 2001); (vi) the molecular abundance across PSCs is set by freeze-out/desorption processes in the core centers and by photodissociation in the outskirts (Pavlyuchenkov et al. 2008; Keto \& Caselli 2008); (vii) PSCs are thermally supercritical cloud cores (Keto \& Caselli 2008), with infall velocities of $\simeq$0.1\,km\,s$^{-1}$ peaking at $\simeq$1000\,AU from the core center (Keto \& Caselli 2010).

High angular resolution observations (probing linear sizes of $\simeq$800\,AU) using the Very Large Array (VLA) and the Plateau de Bure Interferometer (PdBI) have shown that the specific angular momentum of the PSC L1544 drops from $\simeq$5$\times$10$^{21}$\,cm$^2$\,s$^{-1}$ at sizes of 0.04 (traced by NH$_3$) to $\simeq$3.5$\times$10$^{20}$\,cm$^2$\,s$^{-1}$ at sizes of 0.001\,pc (traced by NH$_2$D; Crapsi et al. 2007). While the former value is typical of starless cores (e.g. Ohashi et al. 1999), the smaller obtained toward the core nucleus resembles values measured toward protostellar envelopes and it is consistent with what found in more evolved objects (Belloche et al. 2002). ALMA will allow us to dive into the central few thousands AU and study the physical conditions and kinematics of the core nucleus, the future protostar and protoplatenary disk cradle. 

Figure 1 shows a schematic summary of the physical and chemical structure of a representative PSC based on recent observational and theoretical work. 
The panel is divided in five regions with different gray scales, representing various chemical zones: (i) the "complete depletion" zone ($r$ $\le$ 1000\,AU, $A_{\rm V}$ $\ge$ 60\,mag), not yet discovered but predicted by chemical models (e.g. Walmsley et al. 2004). Here, the main tracers of the gas are expected to be the ortho-H$_2$D$^+$(1$_{10}$-1$_{11}$) and para-D$_2$H$^+$(1$_{10}$-1$_{01}$), both observable with ALMA. (ii) The "depletion zone" (1000 $<$ $r$ $\le$ 5,000\,AU, $A_{\rm V}$ $\ge$ 20\,mag), where the peak of deuterated molecules such as N$_2$D$^+$, D$_2$CO and deuterated ammonia are found (e.g. Caselli et al. 2002a; Bacmann et al. 2003; Crapsi et al. 2007). Other non-deuterated molecules more resistant than CO to the freeze-out have also been observed in the same region: CN, HCN, HNC (Hily-Blant et al. 2010, see also Tafalla et al. 2006). (iii) The "CO freeze-out zone" ($r$ $\le$ 7,000\,AU, $A_{\rm V}$ $\ge$ 5\,mag) where the majority of CO molecules ($>$99\%, Caselli et al. 1999) are locked onto icy dust grain mantles. (iv) The "dark-cloud chemistry zone" (7,000 $<$ $r$ $\le$ 15,000\,AU, $A_{\rm V}$ $\ge$ 4\,mag), traced by CS, CO and higher density tracers chemically linked to CO, such as HCO$^+$. Because carbon atoms are probably not completely locked into CO molecules (due to the presence of interstellar UV photons), carbon-chain species such as CCS and CCH (Tafalla et al. 2006; Padovani et al. 2009) can also be found here. (v)  The "photodissociation region" (PDR; e.g. Kaufman et al. 1999), where FUV (6\,eV $<$ $h \nu$ $<$ 13.6\,eV) photons play a crucial role in the chemistry, dissociating molecules and maintaining most of the carbon and oxygen in atomic form in the gas phase.  The detailed structure of PSCs of course depends on the environment, so that the size and physical/chemical properties of the various PSC layers are expected to change in different molecular cloud complexes or in more massive star forming regions.  Understanding how external conditions affect the structure of PSCs is crucial to shed light on the star formation process in different environments.

\begin{figure}[]
%\vspace*{2.0 cm}
\begin{center}
 \includegraphics[width=3.4in,angle=270,scale=0.9]{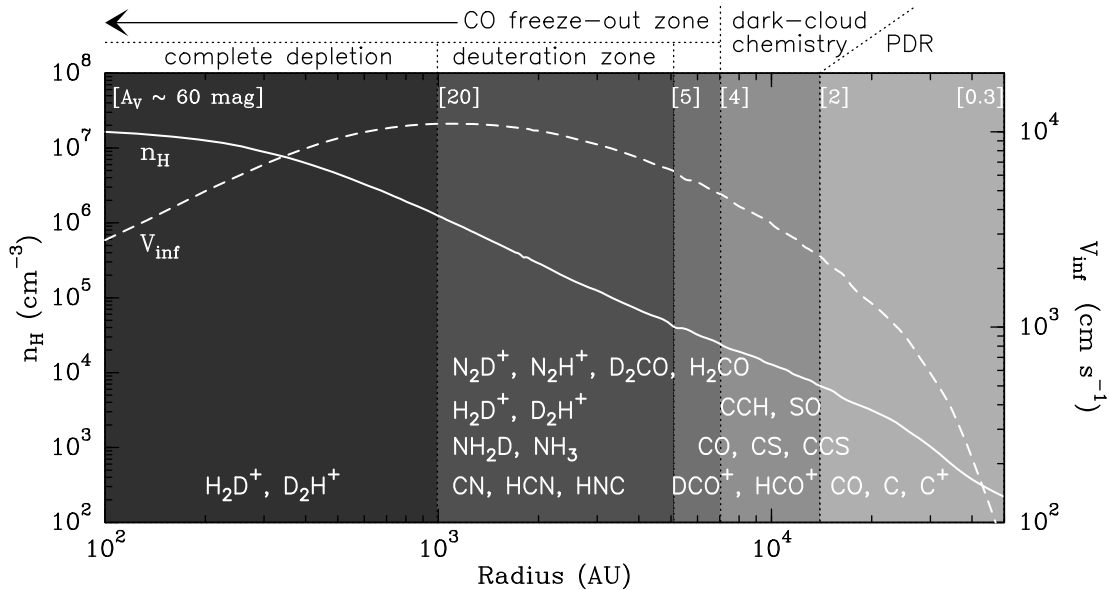} 
% \vspace*{-1.0 cm}
 \caption{Schematic representation of the physical and chemical structure of a PSC in a low-mass star forming region, based on detailed observations and hydrodynamic/radiative transfer modeling of L1544, isolated PSC embedded in the Taurus molecular cloud complex. The white curves represent the hydrogen nuclei volume density ($n_{\rm H}$; solid, left labels) and the infall velocity ($V_{\rm inf}$; dashed, right labels) profiles as derived by Keto \& Caselli (2010). The density is predicted to be 2$\times$10$^{7}$\,cm$^{-3}$ within the central 500\,AU. The infall velocity has a peak of $\simeq$0.1\,km\,s$^{-1}$ around 1,000\,AU.  The panel is divided in five regions with different gray scales, representing various chemical zones (see text).}
   \label{fig1}
\end{center}
\end{figure}

\subsection{The water content}

Oxygen is the third most abundant element in the Universe and it has a crucial role in the chemical and physical structure of molecular clouds, where large fractions are locked into CO and H$_2$O molecules.  Unfortunately, large uncertainties in the chemistry of oxygen are still present due to the difficulty of observing H$_2$O and O in molecular clouds.  In dark ($A_{\rm V}$ $>$ 4\,mag) clouds, H$_2$O is thought to be mostly in solid form (Hollenbach et al. 2009), in agreement with the stringent upper limits on the ortho-H$_2$O(1$_{10}$-1$_{01}$) obtained toward dark clouds and dense cores ($x({\rm H_2O}$ $\equiv$ $N({\rm H_2O})/N({\rm H_2})$ $<$ 3$\times$10$^{-8}$, 8$\times$10$^{-9}$ and 7$\times$10$^{-9}$ toward B68, Oph\,D and Cha-MMS1, respectively; Bergin \& Snell 2002; Klotz et al. 2008).  These limits are however difficult to interpret, given that radiative transfer modeling has shown that line trapping and absorption of the dust continuum can heavily affect the water emission from dense interstellar clouds (Poelman et al. 2007). Moreover, a significant fraction of atomic oxygen can still be maintained in the gas phase, thanks to non-thermal (e.g. cosmic-ray induced) desorption processes (Caselli et al. 2002c), as in fact measured with ISO (Infrared Space Observatory) toward two molecular clouds  (Caux et al. 1999; Vastel et al. 2000). 

The superior angular resolution and sensitivity of the Herschel Space Observatory (Pilbratt et al. 2010) has recently furnished the first column density measure of ortho-H$_2$O toward the PSC L1544 (Caselli et al. 2010, see Figure 2), within the WISH (Water In Star-forming regions with Herschel; van Dishoeck et al 2011) guaranteed time programme. No water line was detected toward the starless core B68 (from which a stringent upper limit of the water abundance was deduced: $x({\rm ortho-H_2O})$ $<$ 1.3$\times$10$^{-9}$). The water detection toward L1544 has been possible because L1544, unlike B68, is centrally concentrated and thus has a relatively bright FIR nucleus (about 5\,Jy at 557\,GHz within the Herschel beam of 40$^{\prime\prime}$). In fact, the ortho-H$_2$O(1$_{10}$-1$_{01}$) line has been detected in {\it absorption} at a 5\,$\sigma$ level (Figure 2). The column density is  $N({\rm ortho-H_2O})$ $\simeq$ 8$\times$10$^{12}$\,cm$^{-2}$. Using the H$_2$ column density deduced by the 1.3\,mm dust continuum emission  (Ward-Thompson et al. 1999; Figure 2) within the Herschel beam, Caselli et al. (2010) found $x({\rm ortho-H_2O})$ $\simeq$ 2$\times$10$^{-10}$ toward the central few thousand AUs, where CO is highly frozen onto dust grains. This value is close to that measured in the envelope of massive young stellar objects (van der Tak et al. 2010). Radiative transfer models of collapsing Bonnor-Ebert spheres with the inclusion of simplified chemistry based on the Hollenbach et al. (2009) model, were able to reproduce the observations assuming a peak abundance of water vapor of $\simeq$10$^{-8}$ at 0.1\,pc from the PSC center. Higher gas phase abundance of water are ruled out because the predicted absorption is too strong. These observations have been recently repeated and the S/N significantly improved so that a more detailed analysis and comparison with chemical models is now possible and will soon be presented (Caselli et al. 2011, in prep.).   

Figure 2 shows the ortho-H$_2$O(1$_{10}$-1$_{01}$) spectrum toward L1544 as presented in Caselli et al. (2010), where a comparison with the CO(1-0) profiles  is also given. It is interesting to see that the water absorption is present in the same velocity range of the CO(1-0) emission line, including the faint CO component redshifted by about 2\,km\,s$^{-1}$ away from the centroid velocity of the L1544 dense gas tracers.  These are the first observations of water vapor in dark clouds and they are crucial to shed light on the oxygen chemistry and the formation of water vapor in quiescent regions.

\begin{figure}[]
% \vspace*{-2.0 cm}
\begin{center}
 \includegraphics[width=3.4in,angle=270,scale=1.0]{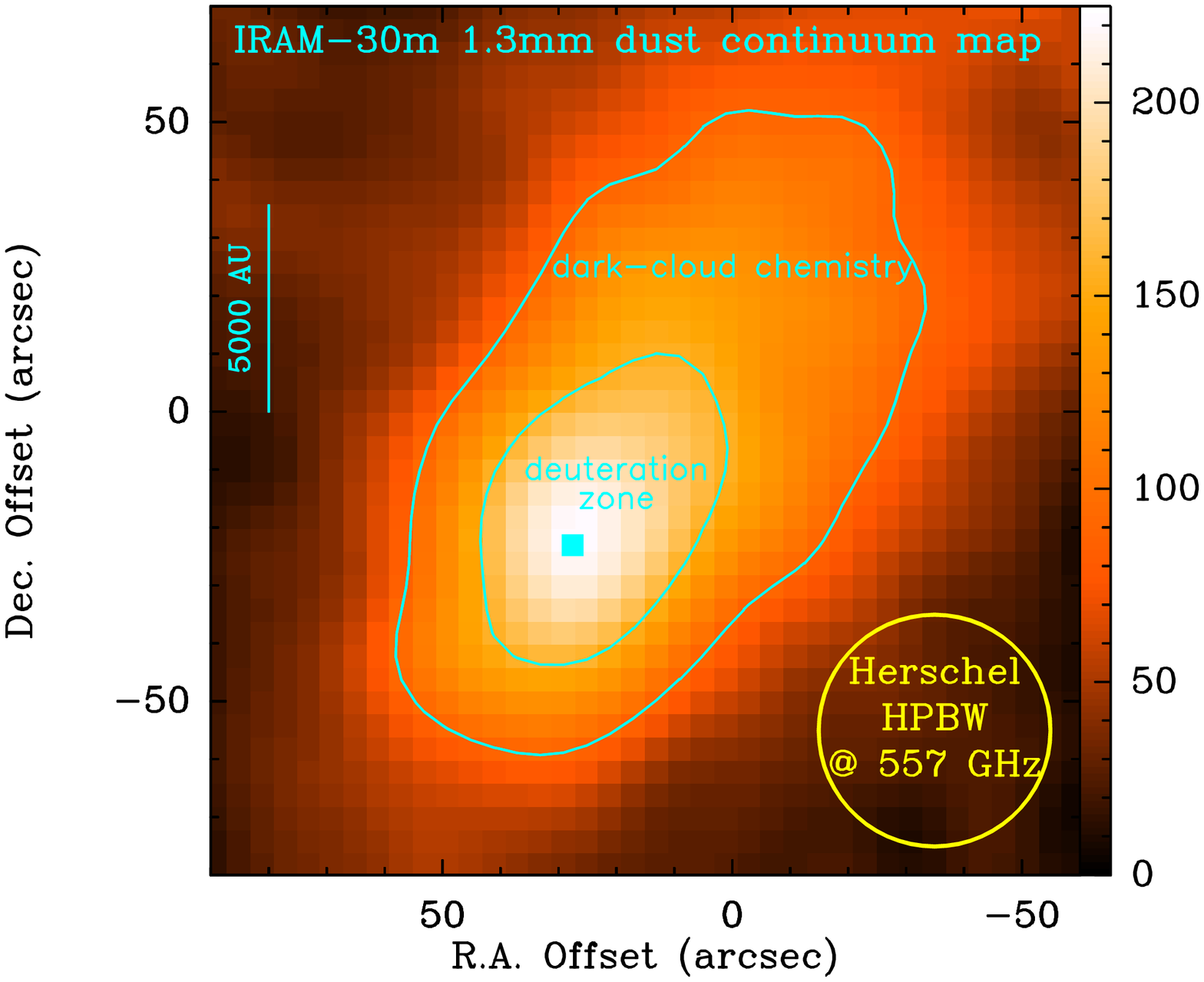} 
  \includegraphics[width=3.4in,angle=270,scale=0.6]{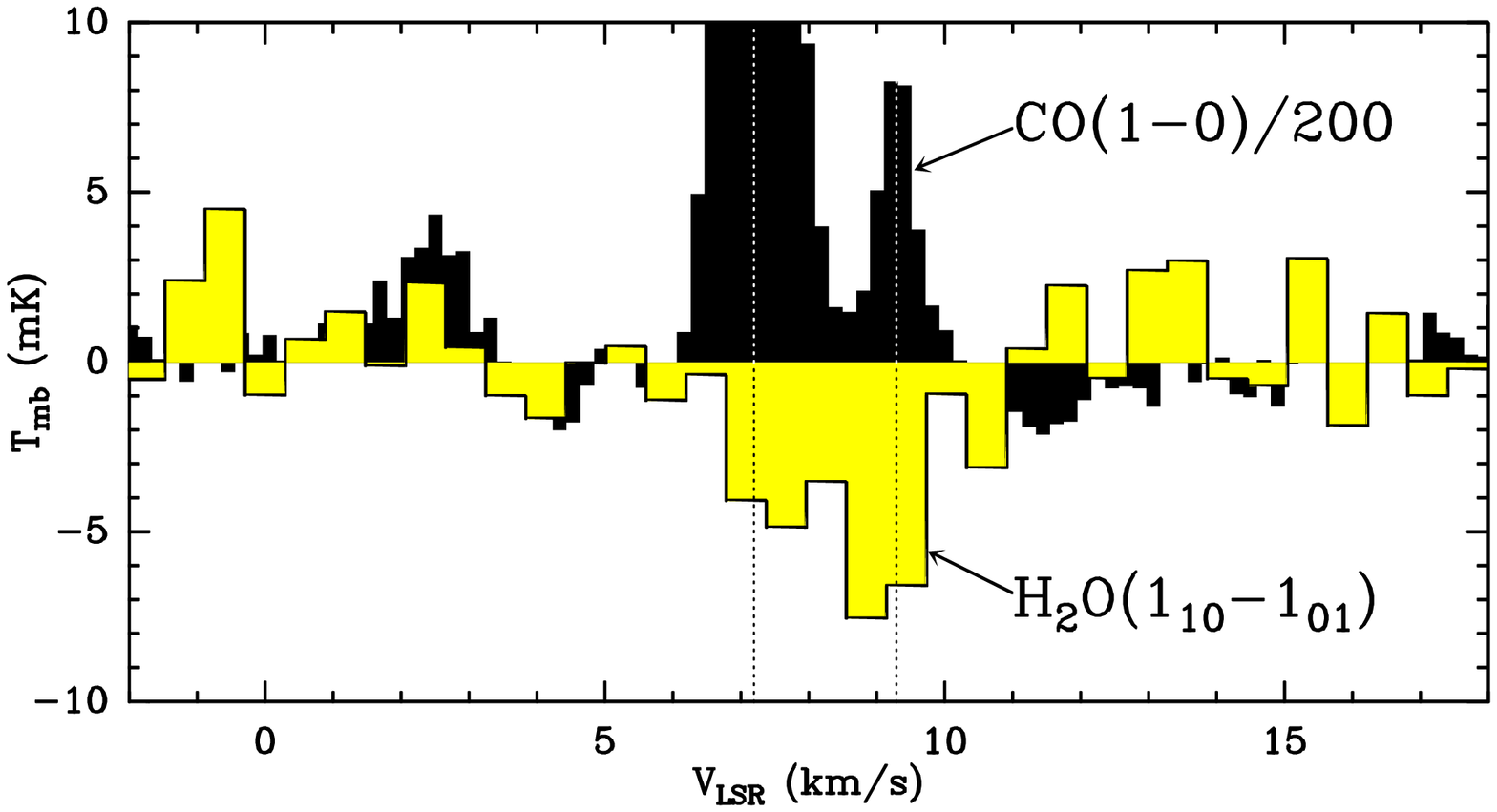} 
 \caption{({\bf Top panel}) The 1.3\,mm dust continuum emission map obtained at the IRAM-30m antenna by Ward-Thompson et al. (1999), smoothed at 22$^{\prime\prime}$ resolution to improve the sensitivity. The wedge on the right is in units of mJy/22$^{\prime\prime}$-beam. The contours represent the 50\% and 70\% of the dust continuum peak (225\,mJy/beam). They approximately coincide with the size of the "dark-cloud chemistry" zone and "deuteration zone", respectively (see Figure 1). The square locates the dust peak position, observed by Herschel  (RA(J2000) = 05$^h$\,04$^m$\,17.21$^s$, Dec(J2000) = 25$^{\deg}$\,10$^{\prime}$\,42.8$^{\prime\prime}$).  The Herschel beam at 557\,GHz is the circle at the bottom-right corner. ({\bf Bottom panel}) The feature in absorption is the ortho-H$_2$O(1$_{10}$-1$_{01}$) line observed against the 11\,mK continuum emission of the dust at the same frequency (from Caselli et al. 2010). The black histogram is the CO(1-0) line (scaled down by a factor of 200) observed with the Five College Radio Astronomy Observatory (FCRAO).}
\label{fig2}
\end{center}
\end{figure}

\section{Massive PSCs}
\label{sec_highmass}

The majority of stars form in clusters within giant molecular clouds (GMCs; e.g. Lada \& Lada 2003), so it is crucial to understand how stellar clusters form. However, the initial conditions in the process of high-mass star and stellar cluster formation are still highly debated. Understanding what causes a particular region of a GMC to attain high densities on the way to forming a star cluster is a major open question in the fields of star and galaxy formation. There are two main classes of theories: spontaneous gravitational instability perhaps regulated by photo-ionization (McKee 1989) or turbulence (Krumholz \& McKee 2005) and triggering by some external process, such as HII regions (Deharveng et al. 2005), supernova blastwaves (Palous et al. 1994), turbulent motions (MacLow \& Klessen 2004) or GMC collisions (Tan 2000).  Detailed studies of the chemical/physical structure and kinematics of massive PSCs (MPSCs) and comparison with the properties of low-mass PSCs (see Sect.\,\ref{sec_lowmass}) are sorely needed. 

\begin{figure}[]
\centering
%\vspace*{-2cm}
\includegraphics[angle=0,scale=.6]{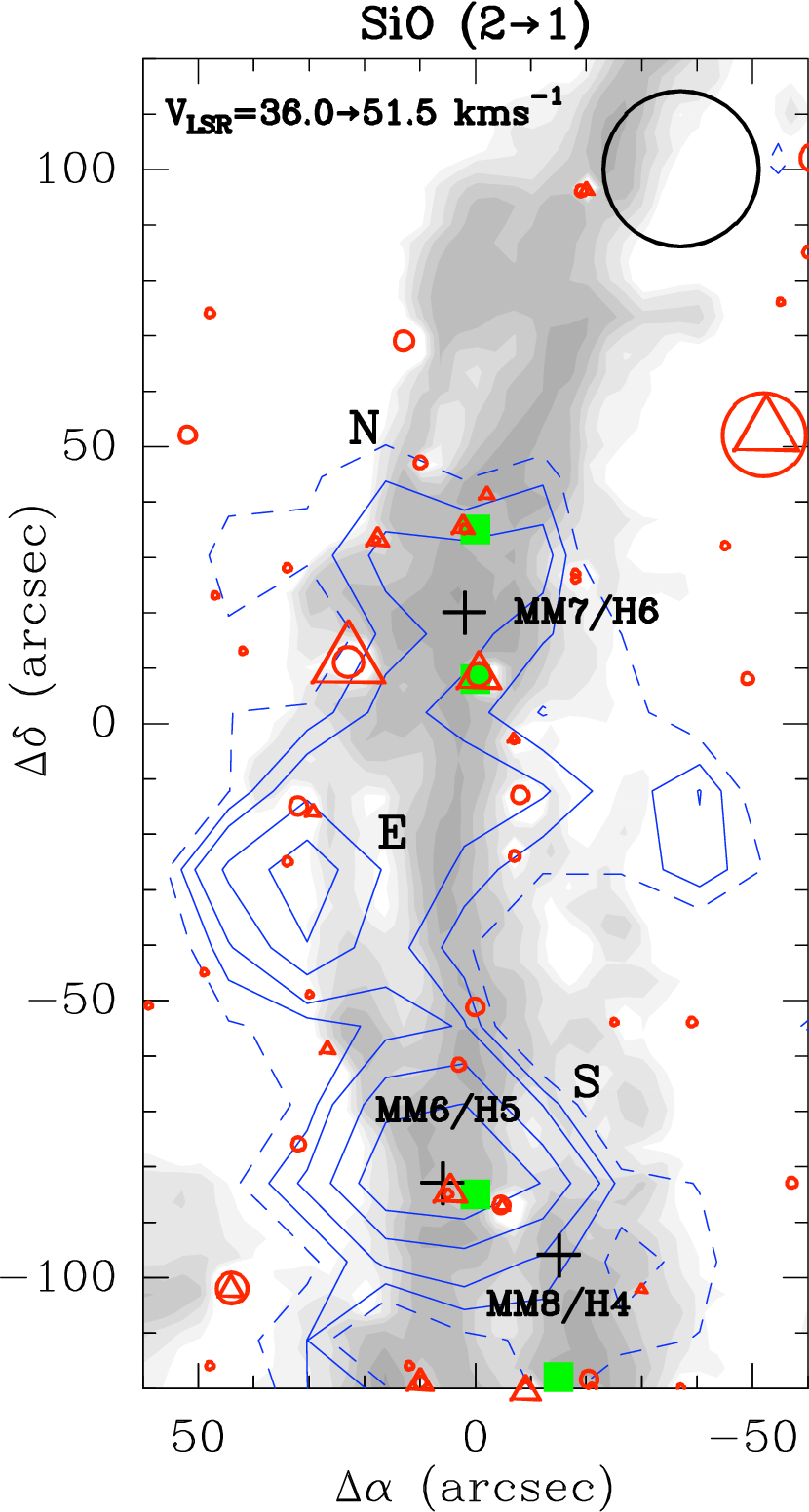}
\includegraphics[angle=0,scale=.38]{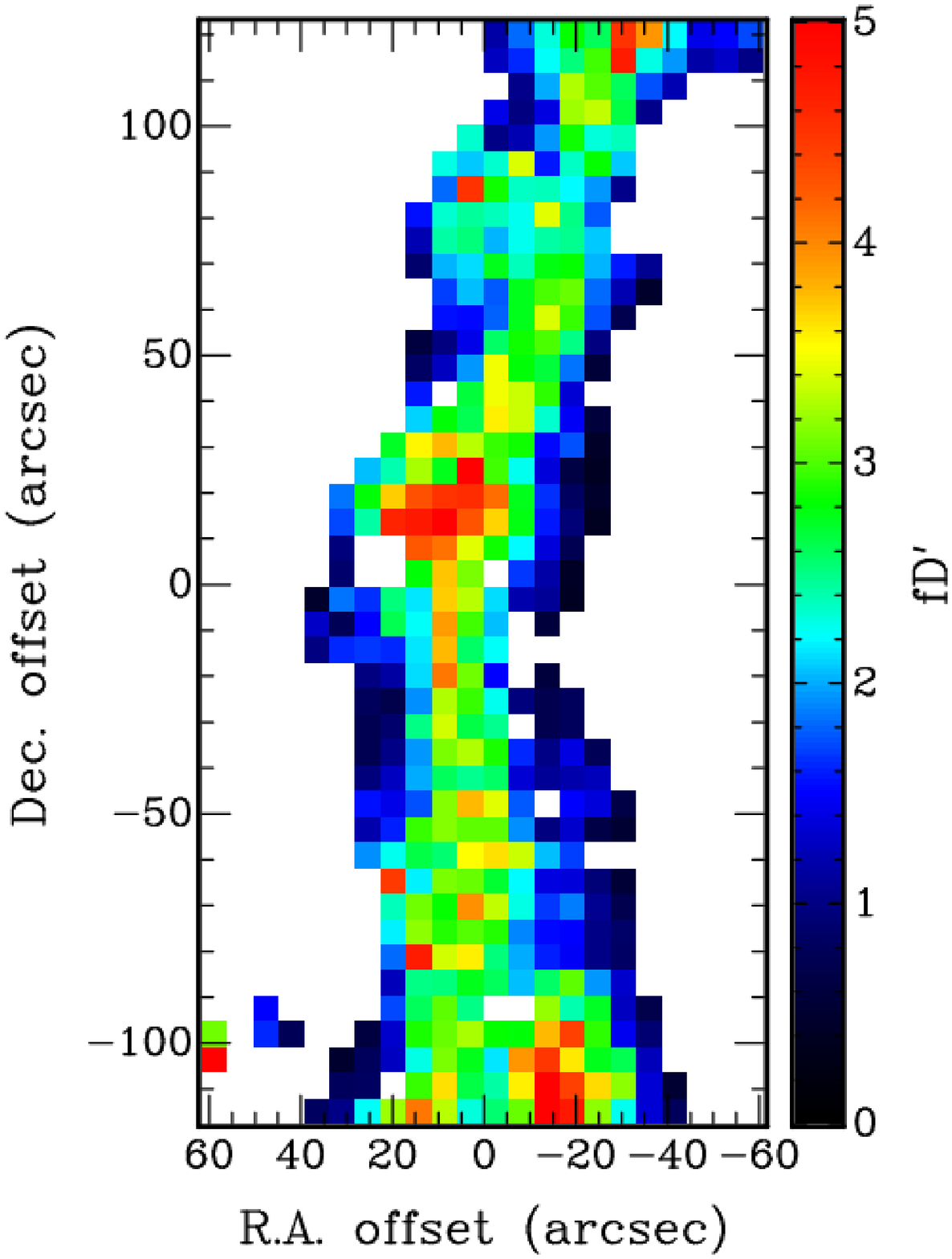}
%\vspace*{-0.3cm}
\caption{\small ({\bf Left panel}) Integrated intensity map of the SiO (2-1) line towards G035.39-00.33 (blue contours), overlapped on the mass surface density map of Butler \& Tan (2009, grey scale). The contour levels of the SiO emission are 10 (2$\sigma$; dashed contour), 15, 20, 30 and 40 mK km s$^{-1}$. For the mass surface density map, contours are 0.014 (2$\sigma$), 0.021, 0.035, 0.049, 0.07, 0.105 and 0.14 g cm$^{-2}$. Crosses indicate the cores found by Rathborne et al. (2006). Red open circles, red open triangles and green squares show the location of the 8- and 24-$\mu$m sources  and 4.5-$\mu$m extended emission, respectively. The marker sizes used for the 8- and 24-$\mu$m sources have been scaled by the source flux.  The SiO (2-1) beam size is plotted at the upper-right corner (adapted from Jimenez-Serra et al. 2010).  ({\bf Right panel}) CO depletion factor map (from Hernandez et al. 2011) of G035.39-00.33. The abundance of CO decreases by factors 3-5 in the highest extinction regions, indicating the presence of cold dust ($T$ $<$ 20\,K) and consequent CO freeze-out.}
\label{fig3}
\end{figure}

The search and study of MPSCs in high-mass star forming regions is however at its infancy, given that it is challenged by the MPSC short time scales, large distances and their proximity to active sites of high-mass star and cluster formation, which can affect their dynamical evolution. For these reasons, the initial conditions in the process of high-mass and stellar cluster formation require the search for relatively quiescent massive clouds 
%(with masses $>$ 870\,M$_{\odot}$($r$/pc)$^{1.33}$, where $r$ is the cloud radius; Kauffmann \& Pillai 2010) 
and then study the embedded stellar population content to identify starless regions, possibly away from the reach of powerful outflows driven by massive young stellar objects (MYSO) embedded in nearby clumps.  The best candidates to host {\it pristine} MPSCs are the so-called infrared dark clouds (IRDCs; e.g. Beuther et al. 2007; Zhang et al. 2009), in particular those with  a relatively large fraction of quiescent gas (i.e. low star formation activity) and filamentary structures, which are predicted to be the initial morphologies of GMCs in several classes of dynamical formation models (e.g. van Loo et al. 2007; Heitsch et al. 2009). 

IRDCs are high extinction regions viewed against the diffuse mid-IR Galactic background (P\'erault et al. 1996; Egan et al. 1998; Peretto \& Fuller 2009).  They have very high-mass surface densities that are similar to known regions of massive star and star cluster formation (Tan 2005). They are cold (temperatures between 8 and 20\,K; Pillai et al. 2006; Peretto et al. 2010; Stamatellos et al. 2010; Devine et al. 2011; Ragan et al. 2011), host clumps with densities $\simeq$10$^4$-10$^6$\,cm$^{-3}$ (Teyssier et al. 2002; Butler \& Tan 2009; Swift 2009; Zhang et al. 2011), show chemical compositions similar to low-mass PSCs (Vasyunina et al. 2011), including large deuterium fractions $R_{\rm D}$ (Pillai et al. 2007; Chen et al. 2010).  In fact, Fontani et al. (2011) have found significantly larger N$_2$D$^+$/N$_2$H$^+$ column density ratios toward a sample of high mass starless cores (HMSCs) embedded in cold IRDCs (between 0.3 and 0.7) than the rest of their sample, which includes HMSCs embedded in more active high-mass star forming regions as well as clumps associated with high-mass protostellar objects and ultra-compact HII (UCHII) regions.   Thus, $R_{\rm D}$ measurements are powerful tools to identify MPSCs among the IRDC starless cores, which are now readily found by Herschel on a galactic scale (e.g. Molinari et al. 2010; Wilcock et al. 2010; Henning et al. 2010).

Detailed mapping and analysis has been recently done (and more it is under way) of a highly filamentary IRDC: G035.39-00.33. This region (i) has one of the most extreme filamentary structure in the Rathborne et al. (2006) sample of IRDCs studied in the mm continuum emission; (ii) shows extended quiescent regions with high-mass surface density and relatively low star formation activity (as can be seen from the lack of a large population of sources in the Spitzer 24\,$\mu$m image; see Figure 3, left panel); (iii) is relatively nearby (D = 2.9 kpc). This IRDC overlaps with Spitzer-GLIMPSE data, and is Òcloud HÓ in the sample of 10 clouds (A-J) studied by Butler \& Tan (2009) using extinction mapping in the 8$\mu$m IRAC band. Hernandez \& Tan (2011) have studied the kinematics of the gas associated with this cloud using $^{13}$CO(1-0) from the Galactic Ring Survey (GRS; Jackson et al. 2006) and found evidence for disturbed kinematics, perhaps indicating the filament has formed very recently. Widespread SiO(2-1) emission has been found along the filament, suggestive of a large scale shock caused by the recent collision of molecular filaments, which may have produced the IRDC (Jimenez-Serra et al. 2010; Figure 3, left panel). Hernandez et al. (2011) have measured signiÞcant levels of CO freeze-out (factor of 5; see Figure 3, right panel), indicative of the presence of cold and dense gas. Henshaw et al. (2011, in prep) found that the main starless core in IRDC is located at the position where different filaments appear to merge, indicating a recent formation via cloud-cloud collision. 
The presence of extended SiO emission and CO freeze-out, plus the evidence of filament merging (Henshaw et al. 2011, in prep.), further suggest that this region is dynamically young, so it is an ideal target where to study the initial conditions in the process of high-mass star and stellar cluster formation.
Higher angular resolution observations are under way to connect the large scale structure of IRDC with the embedded PSCs and clumps.  This will allow us to make quantitative comparisons with models of giant molecular cloud (GMC) formation and evolution (e.g. Heitsch et al. 2009, 2011; Butler et al., in prep.), and achieve a deeper understanding of the physical processes involved in the earliest phases of high-mass and cluster forming regions. 

\vspace{3mm}

\noindent
{\bf Acknowledgements.}
I am grateful to many collaborators \& friends who made all this work possible and very enjoyable, in particular Luca Bizzocchi, Tyler Bourke, Michael Butler, Cecilia Ceccarelli, Antonio Crapsi, Francesco Fontani, Jonathan Henshaw, Audra Hernandez, Itaskun Jim\'enez-Serra,  Eric Keto, Phil Myers, Jaime Pineda, Scott Schnee, Mario Tafalla, Jonathan Tan, Floris van der Tak, Charlotte Vastel, Malcolm Walmsley.  A special thank to Audra for her help with Figure 3, and to Mario, Jonathan and Malcolm for their careful reading of the manuscript and important suggestions.

\end{document}